\documentclass[prl,showpacs,aps,twocolumn]{revtex4}
\usepackage{amsmath}
\usepackage{epsfig}
\begin{document}

\bigskip
\title{Microscopic theory of the two-proton radioactivity}

\author
{J. Rotureau$^{\dagger}$, J. Oko{\l}owicz$^{\dagger \star}$ and M. P{\l}oszajczak $^{\dagger}$}
 
\affiliation{$^{\dagger}$
Grand Acc\'el\'erateur National d'Ions Lourds (GANIL), CEA/DSM - CNRS/IN2P3,
BP 55027, F-14076 Caen Cedex, France \\
and \\  $^{\star}$
Institute of Nuclear Physics, Radzikowskiego 152, PL-31342 Krak\'ow, Poland }
 
\date{\today}
 
\begin{abstract}
\parbox{14cm}{\rm 
We formulate theory of the two-proton radioactivity based on the real-energy continuum shell model. This microscopic approach is applied to describe the two-proton decay of the $1^-_2$ state in $^{18}$Ne.}
 
\end{abstract}
\bigskip 
\maketitle

One of the main frontiers of the nuclear many-body problem is the structure
of  weakly bound and unbound nuclei with extreme neutron-to-proton ratios. Weakly bound states or resonances cannot be reliably calculated in the closed quantum system formalism. For bound states, there appears a virtual scattering into the continuum phase space which affect the effective 
nucleon-nucleon interaction. For unbound states, the continuum structure appears explicitly in the properties of those states.  Mathematical formulation within the Hilbert space of nuclear states
embedded in the continuum of decay channels goes back to Feshbach
\cite{Fesh}. A unified description of nuclear structure and nuclear
reaction aspects is more complicated and became possible in
realistic situations only recently in the framework of the Shell Model
Embedded in the Continuum (SMEC) \cite{smec1,Oko03}, which is based on the completeness of a one-particle basis consisting of bound orbits and a real-energy  continuum.   The single-particle (s.p.) resonances have to be regularized: they are included in a discrete part of the spectrum
after removing the scattering tails which are fully incorporated in the embedding continuum.
The configuration mixing in the valence space (internal mixing) and through the coupling to the scattering continuum (external mixing) is calculated microscopically and the asymptotic states are obtained in the $S$-matrix formalism \cite{smec1,Oko03}. 
The  SMEC contains coupling to one-particle continuum and could not be applied applied for the description of  Borromean systems or the two-nucleon decays. In this Letter, we formulate the SMEC to include couplings to the two-nucleon decay channels and present first application of this new formalism to the two-proton (2p) radioactivity. 

\paragraph*{Two-particle continuum in SMEC.---}

Hilbert space is divided in three subspaces : Q,
P and T. In Q subspace, A nucleons are distributed over (quasi-)bound
single-particle (qbsp) orbits. In P, one nucleon is in the non-resonant continuum and A-1
nucleons occupy qbsp orbits. In T, two nucleons are in the non-resonant 
continuum and (A-2) are in qbsp orbits. The coupling between Q, P and 
T subspaces changes the 'unperturbed' Shell-Model (SM) Hamiltonian ($H_{QQ}$) in Q  into the effective Hamiltonian :
\begin{eqnarray}
\label{eq1}
{\cal H}_{QQ} = H_{QQ} &+& H_{QT}G^{(+)}_{T}(E)H_{TQ} \nonumber \\
&+& \left[ H_{QP} + H_{QT}G_{T}^{(+)}(E)H_{TP}\right] {\tilde {G}}^{(+)}_{P}(E) 
\nonumber \\ &\times&
\left[ H_{PQ} + H_{PT}G^{(+)}_{T}(E)H_{TQ}\right]   ~ ,
\end{eqnarray}
\noindent
where superscript '+' denotes outgoing boundary condition,
${\tilde {G}}^{(+)}_{P}(E) = 
\left[ E^{(+)}-H_{PP}-H_{PT}G^{(+)}_{T}(E)H_{TP}\right]^{-1}$
is the Green's function in P modified by the coupling to T, and 
$G^{(+)}_{T}(E) = \left[ E^{(+)}-H_{TT}\right]^{-1}$ 
is the Green's function in T.  In the above equations, $H_{PP}$ , $H_{TT}$  are the unperturbed Hamiltonians in P, T subspaces, respectively, and  $H_{QP}$, $H_{PQ}$ , $H_{PT}$, $H_{TP}$ 
are the corresponding coupling terms between Q, P, and T subspaces. The second term on the r.h.s. of (\ref{eq1}) describes a di-proton emission, and the third term describes a modification 
due to the mixing of sequential 2p, di-proton and one-proton (1p) decay modes. In the following, we shall discuss two limits of this general process : (i) indirect 2p-emission (sequential 2p-emission is a special case of this limit) $H_{TQ}=H_{QT}=0$, and (ii) direct 2p-emission $H_{TP}=H_{PT}=0$. In both limits, the interference of 2p- and 1p-emissions is taken into account by the external mixing of SM wave functions.

\paragraph*{Indirect 2p-emission.---}
In this limit, the effective Hamiltonian (\ref{eq1}) can be written in a convenient form which separates 1p- and 2p-coupling terms:
\begin{eqnarray}
\label {eq2}
&&{\cal H}_{QQ}=H_{QQ} + H_{QP}G_P^{(+)}(E)H_{PQ}  \\
&+& \left[ H_{QP}{\tilde {G}}_P^{(+)}(E)H_{PT}\right]G^{(+)}_T(E)\left[ H_{TP}G^{(+)}_P(E)H_{PQ}\right]
\nonumber
\end{eqnarray}
, where $G^{(+)}_P(E)=\left[ E^{(+)}-H_{PP}\right]^{-1}$ is the Green's function in P. To calculate the 2p-emission width of a given SM state $\Phi_j^{(A)}$, we begin by diagonalization of first two terms on the r.h.s. of (\ref{eq2}) in the SM basis $\{ \Phi_i^{(A)}\}_{(SM)}$ and obtain new many-body states $\{\Psi_i^{(A)}\}_{(1p)}$  in the parent system which include the configuration mixing due to the coupling to the 1p-continuum.  From a chosen decaying state $\Psi_j^{(A)}$, we go on to calculate the 
width due to the coupling to the 2p-continuum (the last term in (\ref{eq2})). 
In the following, we shall assume that the indirect 2p-emission  is the sequential process,  {\it i.e.} the first emitted proton is a spectator of the second emission. This implies a following identification : $H_{PP} \rightarrow H_{Q'Q'}+{\rm {\hat p}}h_0{\rm {\hat p}}$, $H_{TT} \rightarrow H_{P'P'}+{\rm {\hat p}}h_0{\rm {\hat p}}$, where primed quantities refer to (A-1)-nucleon space, {\it i.e.} in Q${'}$ subspace, (A-1) nucleons are in qbsp orbits and in P${'}$ subspace one nucleon is in the continuum and (A-2) nucleons are in qbsp orbits. $h_0$ is a one-body potential describing an average effect of (A-1) particles on the emitted proton and ${\rm {\hat p}}$ denotes a projector on the one-particle continuum states. With this identification, $H_{PT}$ becomes a coupling between newly defined Q${'}$ and P${'}$ subspaces.

\paragraph*{Direct 2p-emission.---}
The effective Hamiltonian describing a di-proton emission becomes:
\begin{eqnarray}
\label{eq3}
{\cal H}_{QQ}=H_{QQ} + H_{QP}G_P^{(+)}(E)H_{PQ} + H_{QT}G_T^{(+)}(E)H_{TQ} 
\end{eqnarray}
As before, we begin by calculating many-body states $\{\Psi_i^{(A)}\}_{(1p)}$ in the parent nucleus. With the new initial state $\Psi_j^{(A)}$, we calculate the 2p-decay width:  
\begin{eqnarray}
\label{eq4}
\Gamma_{(2p)}&=&
-2 Im\left( \langle \Lambda_j ^{(A)}|H_{QT}G^{+}_{T}(E)H_{TQ} | \Lambda_j^{(A)} \rangle\right) 
\\ \nonumber &= &-2 Im \left(\langle w_j^T|\omega_j^{T,(+)}\rangle  \right) ~ \ ,
\end{eqnarray}
where  $\Lambda_j^{(A)}$ is an intrinsic state wave function of the parent nucleus corresponding to $\Psi_j$, $|w_j^T\rangle = H_{TQ}|\Lambda_i^{(A)}\rangle$ is the source term, 
and $|\omega_j^{T,(+)}\rangle$ is a continuation of $| \Lambda_j^{(A)}\rangle$ in T. In general, the description of a 2p-decay into T subspace requires a formulation for the three-body asymptotic. 
In this case, $\omega_j^{T,(+)}$ is expanded in hyper-spherical harmonics three-body Jacobi coordinate system. The details of the CC formalism with three-body asymptotics in the SMEC will be published elsewhere \cite{future}. In the following, we shall approximate the 2p-decay by the emission of (2p)-cluster and the final-state interaction between the two protons in terms of the $s$-wave phase shift. A similar scenario has been used in SM+$R$-matrix model calculations of di-proton decay \cite{bar0,bar1}. The decay channel is specified by :
$c(U) = \{ {\Lambda}_k^{(A-2)};[J_{(k)}, (\phi_{00}(U),S,L)^{J_{(2p)}}]^J \}$, where ${\Lambda}_k^{(A-2)}$ is the intrinsic state wave function of a daughter nucleus corresponding to a non-spurious SM state $\Phi_k^{(A-2)}$, $J_{(k)}$ is the angular momentum of a daughter system, $\phi_{00}(U)$ is the $0s$ intrinsic state wave function of (2p)-cluster with intrinsic energy $U$ and spin $S=0$, $L$ is the relative angular momentum of a (2p)-cluster and a daughter nucleus, and $J$ is the total angular momentum of a system $[({\rm A-2})\otimes({\rm 2p})]$. The source term in (\ref{eq4}) is expanded in the harmonic oscillator (HO) basis  : 
\begin{eqnarray}
\label{eq5}
w_{j,c(U)}^T(r)&=&\sum_n\left(\frac{A}{A-2}\right)^{(2n+L)/2}u_{n,L}(r)\\
\nonumber &\times&\langle \Phi_k^{(A-2)},\phi_{00},S=0,L,n|H_{TQ}|\Psi_j^{(A)}\rangle ~ \ ,
\end{eqnarray}
where $r$ is the relative distance between a daughter nucleus and a (2p)-cluster and $u_{n,L}$ is the HO wave function. The source in (\ref{eq5}) is independent of the intrinsic energy $U$ of a (2p)-cluster. The continuation of $\Lambda_j^{(A)}$ in $T$ is a solution of  the in-homogeneous coupled-channel (CC) equations :
\begin{eqnarray}
\label{eq6}
(E-H_{TT})|\omega_j^{T,(+)}\rangle=|w_j^T\rangle ~ \ . 
\end{eqnarray}
For $H_{TT}$, we assume that the total system $[({\rm A-2})\otimes ({\rm 2p})]$ can be considered as a two-body system in the average potential $U_0$. Projecting (\ref{eq6}) on the channel $c(U)$ gives :
\begin{eqnarray}
\label{eq7}
\left[ E-(E_{k}+U)-{\hat T}\left( \frac{P^2}{2\mu}+U_0(r)\right){\hat T}\right]\omega_{j,c(U)}^{T,(+)}(r)=w_{j,c(U)}^T(r)  \nonumber
\end{eqnarray}
where $E_{k}$ is the intrinsic energy of a daughter nucleus. $P^2/2\mu$ is the internal kinetic energy of the system $[({\rm A-2})\otimes ({\rm 2p})]$ with the reduced mass $\mu$ and ${\hat T}$ is a projection operator on the subspace of cluster states in the continuum of  $P^2/2\mu+U_0(r)$. The internal energy of the cluster $U$ is distributed according to the density of states  function for the two protons $\rho (U)$ \cite{bar0}.
The di-proton decay width is then :
\begin{eqnarray}
\label{eq8}
\Gamma_{(2p)}=-2 Im \left( \int_0^{Q_{2p}}\langle\omega_{j,c(U)}^{T,(+)}|w_{j,c(U)}^T\rangle\rho(U)dU \right) ~ \ .
\end{eqnarray}

\begin{figure}[h]
\centerline{\includegraphics[height=8cm,angle=-90]{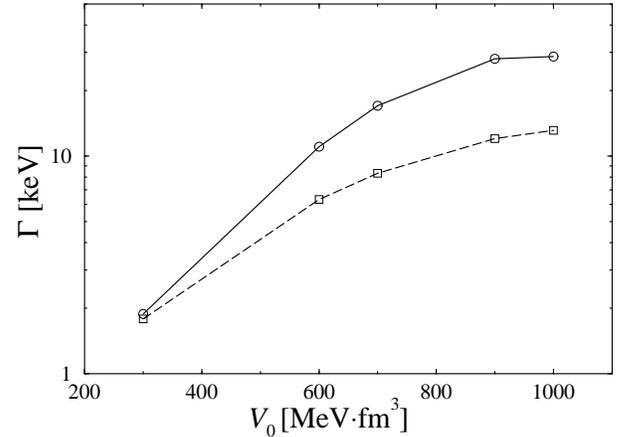}}
\caption{1p-emission width for the decay of $1^-_2$ state in $^{18}$Ne is plotted as a function of the strength $V_0$ of the WB force for the psdfp (solid line) and WBT (dashed line) effective interactions. }
\label{fig1}
\end{figure}

\paragraph*{Two-proton decay of the $1_2^-$ state in $^{18}$Ne.---}

Nuclear decays with three fragments in the final state are very exotic processes. The 2p radioactivity is an example of such a process which can occur for even-Z nuclei beyond the proton drip line : if the
sequential decay is energetically forbidden by pairing correlations, a simultaneous 2p-decay becomes the only possible decay branch. The diproton decay may also be observed in a situation where a 
1p-decay is allowed, as suggested recently in the decay of $1_2^-$ state at 6.15 MeV in $^{18}$Ne \cite{ornl_data}. Here, the decay is dominated by a 1p-decay to weakly bound states $5/2_1^+$ and $1/2_1^+$ in $^{17}$F. No intermediate resonances are available in $^{17}$F which would enable a standard indirect 2p-emission to the ground state (g.s.) of $^{16}$O. The experimental width for the 2p-emission is $21\pm3$ eV assuming a  diproton model, and $57\pm6$ eV assuming a sequential decay model. In describing the decay of $1_2^-$ state in $^{18}$Ne, one has to take into account an influence of the 1p-decay process on the 2p-decay through the $\{5/2^+\}$- and $\{1/2^+\}$-scattering states correlated by weakly bound states $5/2_1^+$ and $1/2_1^+$ of $^{17}$F.  

In SMEC, the radial s.p. wave functions in Q and the scattering wave functions in P are generated by a self-consistent procedure starting with the average potential of Woods-Saxon (WS) type with spin-orbit and Coulomb parts included, and taking into account the residual coupling between Q and P (the procedure of determining the self-consistent potentials in the CC equations and the regularization of s.p. resonances is described in \cite{smec1,Oko03}). This procedure yields new orthonormalized wave functions in Q, P and T and new self-consistent potentials for each many-body state in Q. In this paper, for the effective interaction in $H_{QQ}$ and $H_{Q'Q'}$, we take either WBT Hamiltonian \cite{wbt} or USD Hamiltonian for the $(sd)$-shell \cite{usd}, the KB' interaction for the $(pf)$-shell \cite{kbprim}, and the $G$ matrix \cite{gmatrix} for the cross-shell interaction. The latter interaction is called the psdfp Hamiltonian. Both psdfp- and WBT- Hamiltonians yield the overall correct energies of '$0\hbar\omega$' and '$1\hbar\omega$' states in this mass region. The residual couplings between Q and the embedding continuum is given by the Wigner-Bartlett (WB) force : 
$V_{12}=-V_0[\alpha + (1-\alpha)P_{12}^{\sigma}]\delta(r_1-r_2)$ with $\alpha=0.73$
and the strength $V_0$ which is adjusted to the 1p-decay width of $1_2^-$ state in $^{18}$Ne.

The first two terms on the r.h.s. of (\ref{eq2}) give an effective Hamiltonian describing the 1p-decay and the mixing of SM states due to the Q - P coupling. The reference potential  used in the description of 1p-emission of $1_2^-$ state in $^{18}$Ne (`$^{18}$Ne' parametrization) has the radius  $R_0=3.28$ fm, the surface diffuseness $a=0.58$ fm,  the strength of spin-orbit potential 
${\bar V}_{so}=3.68$ MeV. The depth ${\bar V}_0=-57.62$ MeV of the central part is adjusted to yield 
the s.p. state $0d_{5/2}$ at the experimental 1p-separation energy. $1p_{1/2}, 1p_{3/2}, 0f_{5/2},  0f_{7/2}$ s.p. resonances in this potential are regularized, as described in \cite{smec1,Oko03}, taking $r_{cut}=5$ fm (6 fm) for $0f$ ($1p$) resonances, which corresponds approximately to the top of the barrier. The diffuseness of the resonance cutoff is 1 fm. This reference WS potential \cite{smec1,Oko03} defines energies of s.p. states and yields radial s.p. wave functions for those channels which are {\em not} affected by the continuum coupling. For all other channels, the diagonal potentials in CC equations which include the continuum-coupling correction are constrained to reproduce  reference potential s.p. energies \cite{smec1,Oko03}. 
 Fig. 1 exhibits the 1p decay width of $1_2^-$ state as a function of the strength $V_0$ of the WB coupling to the embedding continuum. The 1p width saturates for large $V_0$ which is a well-known feature of open quantum systems.  The calculated width is smaller than the experimental value $50\pm5$ keV. In the following, we shall take 
$V_0=900$ MeV$\cdot$fm$^3$ for the strength of the coupling between Q, P and Q, T subspaces. 

\begin{figure}[h]
\centerline{\includegraphics[width=8cm]{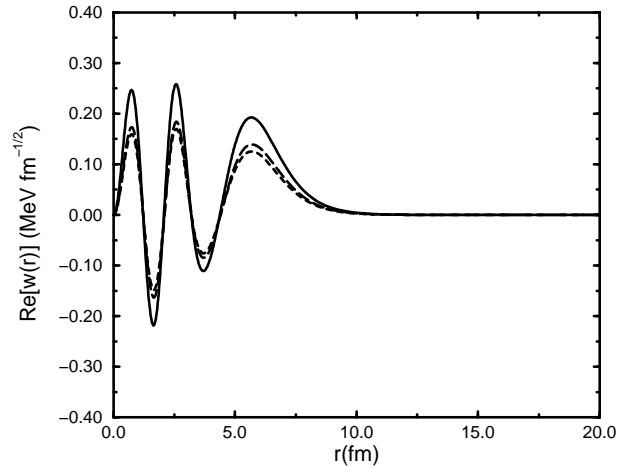}}
\caption{The real part of a di-proton source function for the decay of $1_2^-$ state in $^{18}$Ne. Solid (dashed) line corresponds to the psdfp (WBT) effective interaction. The short-dashed line denotes the source function obtained by neglecting couplings to opened 1p-emission channels. For more details, see the description in the text.}
\label{fig2}
\end{figure}

The sequential 2p-emission to the g.s. of $^{16}$O is possible through the correlated continuum states of $^{17}$F. The theoretical scheme is described above. In a description of the first proton emission into $5/2^+$ and $1/2^+$ continuum of $^{17}$F, we use the `$^{18}$Ne'-reference potential. Reference WS potential used in the description of the emission of the second proton ('$^{17}$F' parametrization)  has the radius  $R_0=3.21$ fm and the surface diffuseness $a=0.58$ fm.  The strength of the spin-orbit potential (${\bar V}_{so}=3.68$ MeV) and the depth  of the central part (${\bar V}_0=-52.46$) MeV are adjusted to reproduce the experimental 1p-separation energies in $^{17}$F for s.p. states $0d_{5/2}$, $1s_{1/2}$. Potential $h_0$ describing an average effect of (A-1) particles on the first emitted proton is given by the self-consistent potential obtained from `$^{18}$Ne' reference potential. 

The width for sequential 2p-emission $\Gamma_{(2p)}^{(seq)}$ depends strongly both on the effective SM interaction and on the coupling to 1p-decay channels. If the coupling to the 1p-scattering continuum is neglected, {\it i.e.} the external mixing of SM states is neglected, then : $\Gamma_{(2p)}^{(seq)}=13.1$ and 38 eV for psdfp and WBT interactions, respectively. The branching ratio $B_{1/2^+}$ for the 2p-decay through the $\{1/2^+\}$-continuum in $^{17}$F is  87\% and 96\% for these two interactions. $\Gamma_{(2p)}^{(seq)}$ and $B_{1/2^+}$ depend on the  total intensity of $(1s0d)(1p0f)$ component in $1_2^-$ state which is 5.8\% and 11.3\% for psdfp and WBT interactions, respectively. 
Including couplings to the 1p-decay channels one finds : $\Gamma_{(2p)}^{(seq)}=88.8$ and 13.6 eV for psdfp and WBT interactions, respectively, {\it i.e.} a bigger 
 2p width is obtained using a psdfp interaction for which the intensity of $(1s0d)(1p0f)$ component in the $1_2^-$ SM state is smaller than for the WBT interaction but the external mixing 
is much stronger than for the WBT interaction and, moreover, interferes constructively. The intensity of the $(1s0d)(1p0f)$ component in  ${1_2^-}_{(1p)}$ SMEC state is now 18.9\% for psdfp interaction and 8.7\% for WBT interaction. Similarly,  the external coupling is reversing the tendency for branching ratio $B_{1/2^+}$ which becomes 93\% and 80\% for psdfp and WBT  interactions, respectively. Clearly, the coupling to opened 1p-decay channels is essential for understanding the sequential 2p-decay of $1_2^-$ state in $^{18}$Ne.

\begin{figure}[h]
\centerline{\includegraphics[width=8cm]{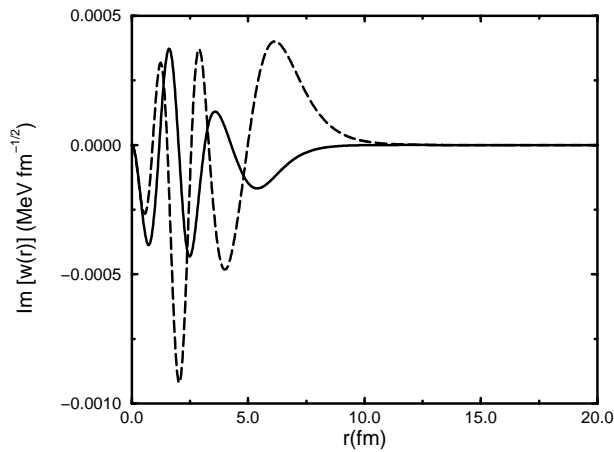}}
\caption{The imaginary part of a di-proton source function for the decay of $1_2^-$ state in $^{18}$Ne. For more details, see the caption of Fig. \ref{fig2} and the description in the text.}
\label{fig3}
\end{figure}

The theoretical scheme used for describing a di-proton decay is given above. Radial s.p. wave functions which enter in the calculation of the source function $w_{j,c}^T(r)$ are generated using  `$^{18}$Ne'-reference potential. In the studied case, the (2p)-cluster is emitted with a relative angular momentum $L=1$. To calculate $\omega_{j,c(U)}^{T,(+)}$, we suppose that the interaction of the (2p)-cluster with a daughter nucleus $^{16}$O is described by an average potential $U_0$ which is a sum of  central WS and Coulomb potentials. Parameters of $U_0$ are deduced from the deuteron scattering data \cite{dae}. The depth of the WS part of $U_0$ (${\bar V}_0=-57.97$ MeV) is adjusted to obtain a $p$-wave resonance for a particle of mass 2$m_p$ and charge $Z=2$ at the energy available for the 2p-decay of $1_2^-$ state in $^{18}$Ne. Fig. \ref{fig2} shows the real part of a di-proton source which for psdfp (the solid line) and WBT (the dashed line) interactions. The short-dashed line in Fig. \ref{fig2} shows results for a psdfp interaction neglecting coupling to 1p-decay channels. All three source functions are qualitatively similar and the Q - P coupling strongly modifies the magnitude of the source. The imaginary part of the source (see Fig. \ref{fig3}) is due to the coupling to 1p-emission channels and strongly differs for psdfp and WBT interactions. The width for a direct 2p-emission $\Gamma_{2p}^{(dir)}$ is 1.89 and 1.01 eV for psdfp and WBT interactions, respectively. Neglecting the coupling to 1p-decay channels, one finds 0.8 and 1.17 eV. Similarly as for the sequential 2p-decay, $\Gamma_{2p}^{(dir)}$ depends strongly on the coupling to the opened 1p-decay channels. 

In summary, we have extended the real-energy continuum shell model (SMEC) to include couplings to the two-particle continuum and applied this new formalism for 1p- and 2p-decays of $1_2^-$ state in $^{18}$Ne. The experimental data are compatible with the sequential 2p-decay through the correlated scattering continuum of $^{17}$F with, possibly, a weak di-proton branch. As compared to the SM+R-matrix calculations \cite{bar1}, we find systematically larger (resp. smaller) two-proton decay width $\Gamma_{(2p)}^{(seq)}$ (resp. $\Gamma_{2p}^{(dir)}$). This difference is mainly due to the  absence of external mixing \cite{Oko03} in SM+R-matrix model \cite{bar0,bar1} and a more realistic description of the emission process in SMEC.
The 2p-decay width is strongly influenced by the interference between external and internal mixings of SM wave functions.
 Strong couplings between 1p- and 2p- emission sectors invalidate a simple picture of direct 2p-decay as a new independent  decay mode, at least if the 1p-decay channels are opened. 

\acknowledgments
We wish to thank F.C. Barker and F. Nowacki for useful discussions.

\end{document}